\documentclass[final,times,5p,xtwocolumn]{elsarticle}

\usepackage{amssymb}
\usepackage{amsmath}
\usepackage{latexsym,epsfig}
\usepackage{graphicx}
\usepackage{hyperref}

\usepackage{amsfonts}
\usepackage{epsfig}
\usepackage{dcolumn}
\usepackage{bm}

\def\to{\rightarrow}

\journal{Physica A}

\begin{document}

\begin{frontmatter}



\title{Maximum Entropy Principle and the Higgs Boson Mass}


\author[a1]{Alexandre Alves}
\author[a2]{Alex G. Dias}
\author[a3]{Roberto da Silva}

\address[a1]{Universidade Federal de S\~ao Paulo, Departamento de Ci\^encias Naturais e da Terra,  Diadema - SP, Brasil}
\address[a2]{Universidade Federal do ABC, Centro de Ci\^encias Naturais e Humanas,  Santo Andr\'e - SP, Brasil}
\address[a3]{Universidade Federal do Rio Grande do Sul, Departamento de F\' isica, Porto Alegre - RS, Brasil}


\begin{abstract}
A successful connection between Higgs boson decays and the Maximum Entropy Principle is presented. Based on the information theory inference approach we determine the Higgs boson mass as $M_H= 125.04\pm 0.25$ GeV, a value fully compatible to the LHC measurement. This is straightforwardly obtained by taking the Higgs boson branching ratios as the target probability distributions of the inference, without any extra assumptions beyond the Standard Model. Yet, the principle can be a powerful tool in the construction of any model affecting the Higgs sector. We give, as an example, the case where the Higgs boson has an extra invisible decay channel within a Higgs portal model. 
\end{abstract}


\end{frontmatter}


\section{Introduction}
\label{intro}

The mechanism of mass generation~\cite{higgs} via spontaneous electroweak symmetry breaking, as realized in the Standard Model (SM)~\cite{weinberg}, is essential to understand the electroweak interactions and has been almost entirely confirmed with the discovery of the Higgs boson and the observation of its couplings to some of the heavier particles of the SM spectrum~\cite{AtlasCMSHiggs}. In the form we know it, electroweak symmetry breaking is able to generate the particle masses but cannot fix their values. An exception is the photon which is massless due the unbroken $U(1)$ gauge symmetry of electromagnetic interactions. Several attempts have been made to predict the Higgs boson mass from extensions of the SM, but until now there is no clue whether the best guesses are meant to represent a higher level of evidence for a more fundamental theory and consequently a deeper understanding of subatomic physics. 

It has been recognized that entropy can be used as an universal tool for statistical inference in the information theory approach~\cite{jaynes,jaynesb,cover}. It is universal in the sense of its wide range of applicability and it is not necessarily tied to physical interpretations, despite the Gibbs-Shannon entropy is equivalent to the usual thermodynamical entropy~\cite{jaynes}.

In this Letter we show that the Maximum Entropy Principle (MEP) is a powerful inference tool which provides us with the most accurate theoretical determination to date of the Higgs mass. The principle also naturally leads us to assign a theoretical probability density function (PDF) for the Higgs boson mass parameter. It does not assume any beyond the Standard Model hypothesis nor untested physical principles, rather it does only assume the present status of physical knowledge encoded in SM. 

Based on this successful mass parameter inference, we propose using MEP in order to furnish further pieces of evidence to favor or not a given theory affecting the Higgs sector. According to this proposal, any candidate model is required to maximize a physically well motivated entropy associated to the Higgs bosons decays to all available channels for a given mass. As an illustration of the usefulness of the proposed inference method, we investigate a Higgs-portal scalar dark matter model. 

It has been firstly observed in Ref.~\cite{david} that the product of the SM Higgs branching ratios is a distribution with a peak close to the experimentally measured value of the Higgs boson mass. Moreover, the author also mentions a possible connection to this observation with entropy arguments in the Higgs decay process, and yielding a potential mean to constrain theoretical extensions of the SM. It was also suggested  in Ref.~\cite{hmaxphotons} that some fundamental principle might be responsible for the maximization of Higgs branching ratio to photons, a fact that could be used for constraining new physics as well. We shall see that all those observations are consequences of MEP.

Our findings suggest further that entropy production in Higgs decays to massless quanta, namely, to photons and gluons, is maximized nearly the same point of the entropy from all channels. This places the system of a large number of Higgs bosons among the most fundamental ones where it is possible to observe MEP in action. 

\section{MEP inference on the Higgs boson mass}
\label{sec1}

Maximum-entropy distributions are the best estimates that can be made from partial knowledge about the PDF of a given set of random variables. The goal of MEP inference is to determine the least-biased probability density function involved in the evolution of a given system from the computation of the Gibbs-Shannon entropy measure where the partial knowledge is coded as a list of constraints.  It is universal in its scope and is known, for example, to lead to the same computation rules of the statistical mechanics, without any further assumptions beyond the usual laws of mechanics (quantum or classical)~\cite{jaynes}. 

Suppose, however, we can compute a parametric family of PDFs from first principles (as the Higgs branching ratios). In this situation, the partial knowledge reflects all the prior information available encoded in the model (the SM, for example) from which one calculates those PDFs. The correspondence between MEP and the Maximum Likelihood Estimate (MLE) is particularly crucial for the inference of the PDF parameters~\cite{cover}.

We are going to show that a physically well motivated entropy measure for an ensemble of Higgs bosons naturally leads us to a MLE for the Higgs mass parameter and the assignment of a PDF for the Higgs mass as well. For that purpose, consider an ensemble of $N$ non-interacting Higgs bosons which decay to SM particles according to their respective branching ratios $BR_i$, $i=\gamma\gamma,\; gg,\; Z\gamma,\; q\bar{q},\; \ell^+\ell^-$, $ WW^*,\; ZZ^*$, where $q=u,d,s,c,b,t$ and $\ell =e,\mu,\tau$. There are 14 primary SM decay modes.

The probability that a system evolves from the $N$ Higgs bosons to a bath of final state SM particles due $m$ energetically accessible decay modes is given by a multinomial distribution 
\begin{equation}
P(\{n_k\}_{k=1}^m)\equiv \frac{N!}{n_1!\cdots n_m!}\prod_{k=1}^{m} (p_k)^{n_k}
\label{a1}
\end{equation}
where 
\begin{equation}
p_k(M_H|\pmb{\theta})\equiv BR_k(M_H,\pmb{\theta})=\frac{\Gamma_k(M_H,\pmb{\theta})}{\Gamma_H(M_H,\pmb{\theta})}
\end{equation}
are the Higgs branching ratios as a function of the Higgs boson mass $M_H$ and the remaining SM parameters $\pmb{\theta}$ related to the electroweak symmetry breaking. The partial decay width to the $k$-th primary channel is denoted as $\Gamma_k$, and $\Gamma_H$ is the total width of the Higgs boson given by the sum $\Gamma_H=\sum_{k=1}^m \Gamma_k$. Form this definition, $\sum_{k=1}^{m} p_k=1$, and $\sum_{k=1}^{m} n_k=N$ where $n_k$ is the number of Higgs bosons decaying to the $k$th mode.

The partial decay widths of the Higgs boson can be easily computed at leading order in the Standard Model, however, a precise prediction must include higher order corrections of the perturbative series. At leading order, the partial widths are given by
\begin{eqnarray}
\Gamma_{f\bar{f}} &=& \frac{N_c G_F m_f^2 M_H}{4\pi\sqrt{2}}\beta_f^3\Theta(M_H-2m_f)\\
\Gamma_{VV} &=& \frac{\delta_V G_F M_H^3}{16\pi\sqrt{2}}\beta_V\left[1-\frac{4m_V^2}{M_H^2}+\frac{3}{4}\left(\frac{4m_V^2}{M_H^2}\right)^2\right]\nonumber \\
&\times& \Theta(M_H-2m_V) \\
\Gamma_{\gamma\gamma} &=& \frac{G_F m_Z^2}{2\pi\sqrt{2}M_H}F_{\gamma\gamma}(M_H,\pmb{\theta}) \\
\Gamma_{gg} &=& \frac{\alpha_s^2 G_F M_H^3}{36\pi^3\sqrt{2}}F_{gg}(M_H,\pmb{\theta})\\
\Gamma_{Z\gamma} &=& \frac{G_F m_Z^4}{\pi\sqrt{2}M_H}\beta_{Z\gamma}^2F_{Z\gamma}(M_H,\pmb{\theta})\Theta(M_H-m_Z)
\end{eqnarray}
Here, $N_c=3(1)$ for quarks(leptons), $\delta_V=2(1)$ for $W^\pm(Z)$, $m_f$ is the fermion mass, $m_V$ a $W$ or $Z$ mass and $G_F$ is the Fermi constant. The velocities $\beta$ are defined as $\beta_f=\sqrt{1-4m^2_f/M_H^2}$, $\beta_V=\sqrt{1-4m^2_V/M_H^2}$, and $\beta_{Z\gamma}=\sqrt{1-m^2_Z/M_H^2}$, moreover $\Theta(x)=1$ if $x>0$, and zero otherwise. Higher order corrections and the loop functions of the partial widths to photons, gluons and $Z$+photon denoted as $F_{\gamma\gamma}$, $F_{\gamma\gamma}$, and $F_{Z\gamma}$, respectively, in the formulas above, can be found in Ref.~\cite{djouadi} and references therein.

We point out that higher order corrections include, beyond the leading $1\to 2$ Born amplitude, $1\to 2$ loop amplitudes, and $1\to 3$ amplitudes involving additional real emissions. This is the only way to consistently compute finite higher order corrections after renormalizing ultraviolet divergences.

The total entropy is now given by the sum of the multinomial distribution of $N$ Higgs bosons decaying to each possible partition of $n_1$ particles of type 1, $n_2$ of type 2, and so on until the $m$-th mode
\begin{eqnarray}
S_N &=& \sum_{\{n\}}^N-P(\{n_k\}_{k=1}^m)\nonumber\; \ln\left[P(\{n_k\}_{k=1}^m)\right]\nonumber\\
&=& \langle -\ln(P)\rangle
\label{a2}
\end{eqnarray}
where $\sum_{\{n\}}^N (\bullet)\equiv\sum_{n_1=0}^N\cdots\sum_{n_m=0}^N (\bullet)\times\delta\left(N-\sum_{i=1}^m n_i\right)$. 

The number of possible configurations involved in the sum of Eq.~(\ref{a2}) is huge for large $N$. Recently, an asymptotic formula up to order $1/N$ has been derived~\cite{jacek} and is given right below at Eq.~(\ref{a3})
\begin{eqnarray}
S_N &=& \frac{1}{2}\ln\left((2\pi N e)^{m-1}\prod_{k=1}^m p_k\right)\nonumber \\
& + & \frac{1}{12N}\left(3m-2-\sum_{k=1}^m\frac{1}{p_k}\right)+{\cal O}\left(\frac{1}{N^2}\right)
\label{a3}
\end{eqnarray}

We checked the formula up to $N=10^9$ using an importance sampling Monte Carlo technique. In practice, for $N\gtrsim 10^7$, the ${\cal O}(1/N)$ term is negligible for the SM Higgs branching ratios $p_k$.

An important remark about $S_N$ for the multinomial distribution is due.  Calculating the entropy of a system with elementary states grouped into macrostates, with consequent loss of information, is the essence of coarse-graining \cite{lesne}. The physically motivated entropy $S_N$ gathers each type of decay into subgroups counting their occurrence numbers and nothing more. Any additional information about their identities is lost. This way, $S_N$ naturally comprises the concept of irreversibility, in obeisance to the second law of thermodynamics~\cite{kandrup}. 

The most important feature of the asymptotic formula of Eq.~(\ref{a3}) is the term involving the product of probabilities $p_k$ in the thermodynamic limit $N\rightarrow\infty$. This is, up to a normalization factor, a joint PDF to observe a fraction $p_1$ of decays of type 1, and so on until a fraction $p_m$ to observe decays of type $m$. With fixed probabilities, the normalized product is the likelihood function of the parameter $M_H$. This induces the assignment of the normalized product $\prod_{k=1}^m p_k(M_H|\pmb{\theta})$ to a theoretical Higgs mass PDF. Moreover, requiring that the Higgs mass parameter maximizes $S_N$ in the thermodynamic limit is equivalent to a MLE of the Higgs boson mass. In this respect, the physically motivated multinomial distribution is fundamental, once the asymptotic limit of its the entropy  (Eq. \ref{a3}) equals the logarithm of the likelihood function $\prod_{k=1}^m p_k(M_H|{\pmb\theta})$.

A maximum of $S_N$ (a concave function in $M_H$), in the thermodynamic limit, is a solution of the equation
\begin{equation}
\lim_{N\rightarrow\infty}\frac{\partial S_N}{\partial M_H} = \frac{\partial S_\infty}{\partial M_H}=0
\label{a4a}
\end{equation}
where 
\begin{equation}
 S_\infty(M_H|\pmb{\theta})\equiv \ln\left(\prod_{k=1}^m p_k(M_H|\pmb{\theta})\right)
\label{a4b}
\end{equation}

 The mass parameter that maximizes $S_\infty(M_H|\pmb{\theta})$, a solution of Eq.~(\ref{a4a}) that we call $\hat{M}_H$, is distributed according to some PDF $P_H(\hat{M}_H)$ which is obtained after marginalizing over all $\pmb{\theta}$ taking into account their current experimental uncertainties. This should not be confused with the theoretical Higgs mass PDF that Eq.~(\ref{a3}) suggests. We calculate $P_H$ just to access the theoretical error on the inferred $\hat{M}_H$ from uncertainties on the remaining SM parameters.

We adopted the SM parameters values, $\pmb{\theta}_{SM}$, recommended by the Higgs Working Group~\cite{hwg} and calculated the SM branching ratios with \texttt{HDECAY}~\cite{hdecay} taking into account all the leading EW and QCD corrections available. We do not take double off-shell top quark decays into account as its contribution is expected to be a negligible next-to-leading order correction to $W^+W^-$ production in the mass region of interest~\cite{djouadi}, so we effectively have $m=13$ decay modes. We checked our results against those quoted in the Higgs Working Group report within less than 1\% for the whole mass range from 10 GeV to 1 TeV. The $P_H$ PDF was estimated by randomly choosing 2000 points in the SM parameter space, calculating $\hat{M}_H$ for each point, and fitting the resulting histogram to a gaussian function. 

\begin{center}
\begin{figure}
\includegraphics[scale=0.55]{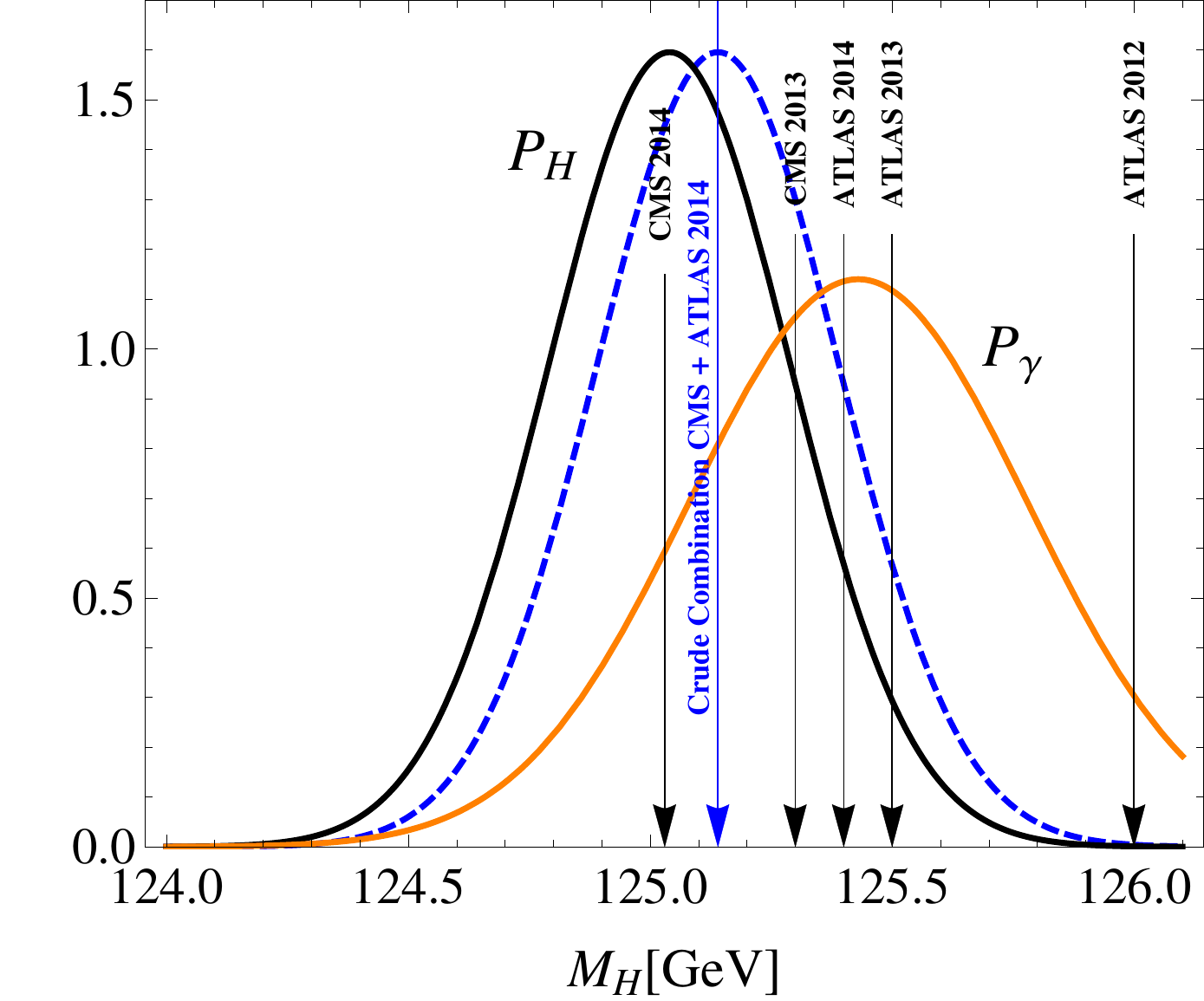}
\caption{
Theoretical and experimental Higgs mass distributions. The solid black line represents the mass distribution obtained from the maximum of the Higgs entropy marginalized over $\pmb{\theta}$, $P_H(\hat{M}_H)$. The dashed line is a crude combination of the most recent ATLAS+CMS results on the Higgs mass. The vertical lines show the experimental evolution of $M_H$ towards the theoretical determination. The solid orange line is the Higgs mass PDF obtained from the maximum of the photons decay entropy also marginalized over $\pmb{\theta}$, $P_\gamma(\hat{M}_H)$.}
\label{fig1}
\end{figure}
\end{center}

After marginalizing over $\pmb{\theta}$, 
\begin{equation}
P_H(\hat{M}_H)=\int_{\Omega_{SM}} P_H(\hat{M}_H|\pmb{\theta})d\pmb{\theta}
\end{equation}
where $\Omega_{SM}$ represents the SM ameters space, we obtained the results shown in Fig.~(\ref{fig1}). The solid black line ($P_H$) is the distribution of the masses that lead to a maximum of $S_\infty$ of Eq.~(\ref{a4b}) while the dashed blue line represents the combination of the most recent CMS and ATLAS measurement of the Higgs mass. The theoretical determination based on MEP inference is
\begin{equation}
\hat{M}_H = (125.04\pm 0.25)\; \hbox{GeV}
\label{a6}
\end{equation}
to be compared with the latest Higgs mass measurements from CMS~\cite{cms14} and ATLAS~\cite{atlas14}
\begin{eqnarray}
 M_H &=& (125.03\pm 0.30)\; \hbox{GeV} \;\;\;\; \hbox{CMS}\nonumber \\
 M_H &=& (125.36\pm 0.41)\; \hbox{GeV} \;\;\;\; \hbox{ATLAS}
\end{eqnarray}
 A crude weighted-averaged combination of both results is $M_H=(125.14\pm 0.24)\; \hbox{GeV}$.
%

In Fig.~(\ref{fig1}) we also show the evolution of the measured Higgs mass since its discovery in 2012 until its most recent determination. We observe an steady decrease in the value of the mass as  the measurements get more accurate converging to the theoretical MEP determination. As far as we know, this is the first time an  underlying principle is used to determine accurately the Higgs boson mass, without  assuming any beyond the Standard Model scenario or untested physical assumptions. 

As discussed above, we can interpret the joint PDF given by the normalized product of branching ratios as a theoretical Higgs mass PDF. What kind of distribution should we expect? Under weak conditions, which we checked that all the branching ratios fulfill, a normalized likelihood function $L$ of a parameter $\xi$ asymptotically converges to the normal distribution ${\cal N}\left(\hat{\xi},1/\sqrt{-L^{\prime\prime}(\hat{\xi})}\right)$ in the mass rehion around the maximum $\hat{\xi}$ in the limit of a large number of PDFs in the product~\cite{fraser}. The mean and the standard deviation (s.d.) of the normalized product of the Higgs branching ratios is given by $\hat{M}_H=125.04$ GeV and $6.13$ GeV, respectively. 

\begin{center}
\begin{figure}
\includegraphics[scale=0.6]{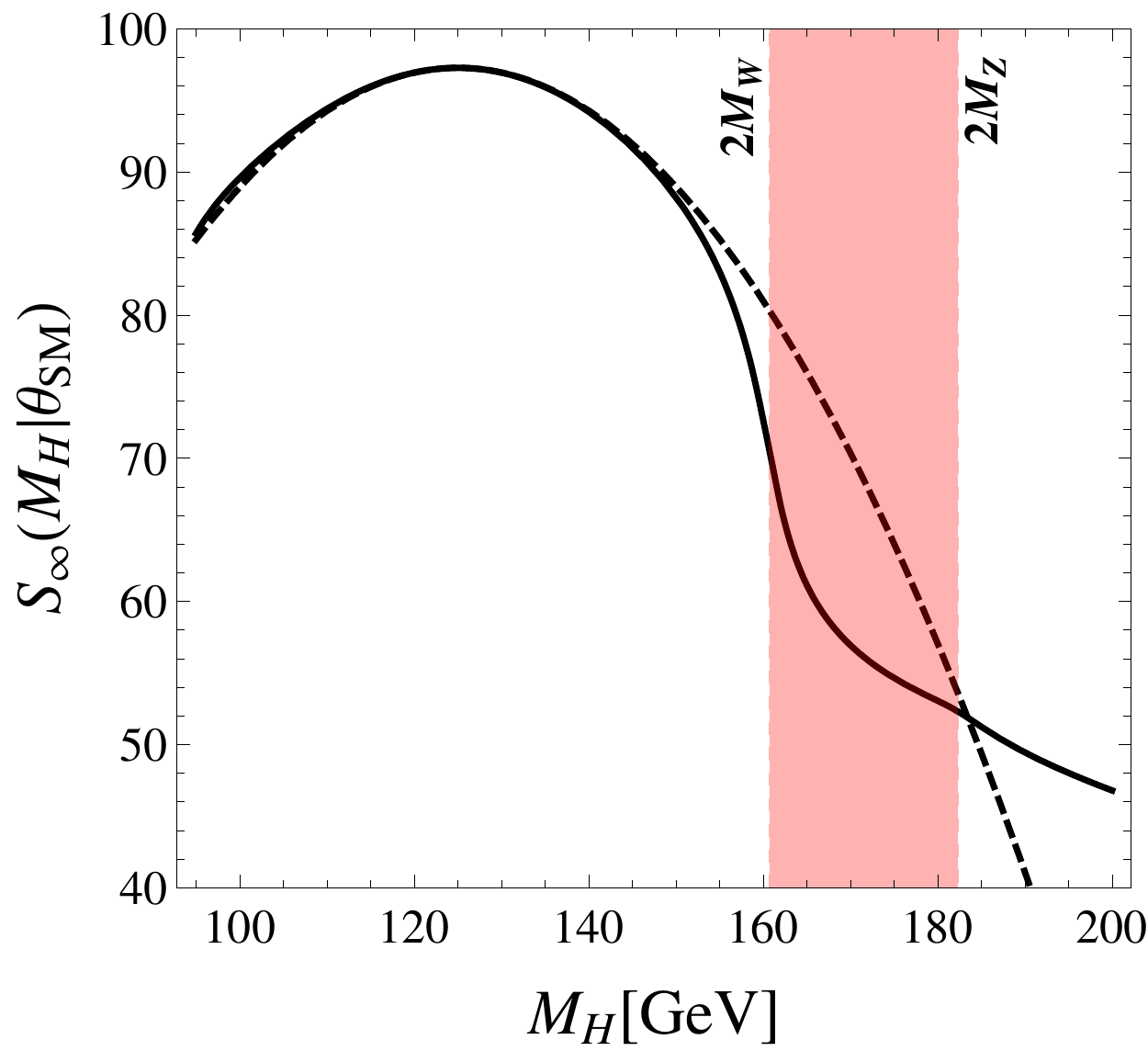}
\caption{The mass dependent term of the multinomial entropy in the thermodynamic limit, Eq.~(\ref{a4b}), as a function of the Higgs mass. The red shaded area emphasizes the threshold region of $W$ and $Z$ pair production where $S_\infty$ drops fast. The dashed line shows a gaussian function with mean $125.04$ and s.d. $6.13$.}
\label{fig2}
\end{figure}
\end{center}

 In fact, the reduced number of branching ratios in the product leads to a strong departure from the gaussian regime as the Higgs mass gets away from the point of maximum as we can see in Fig.~(\ref{fig2}). Furthermore, $S_\infty$, the mass dependent term of the multinomial entropy in the thermodynamic limit of Eq.~(\ref{a4b}), shows the remarkable feature of a fast decreasing in its magnitude at the production thresholds of $W$ and $Z$ pairs as displayed in the shaded red area of Fig.~(\ref{fig2}). This can be understood in terms of an energy reorganization. In the mass region before the $WW$ and $ZZ$ thresholds, the energy condensed in the Higgs mass is spread away by relativistic light particles, by the way, $b$-quarks are the heavier decay products of the Higgs in the region $M_H< 2M_W$. If the Higgs mass was large enough to produce two $W$ or $Z$ on-shell, the branching ratios to $WW$ and $ZZ$ would rise steeply and dominate the decays. In that case, the energy stored in the Higgs mass would be just reorganized in their masses making a rapid fall of the entropy.

 The nearly gaussian shape of $\prod_{k=1}^m p_k(M_H)$ near the maximum was firstly observed in Ref.~\cite{david} with very similar mean and standard deviation (see dashed line in Fig.~(\ref{fig2})). It was argued that the maximum of this distribution (in the sense of MLE), which is close to the experimental mass value, could be due some kind of entropy argument and that the Higgs mass is placed in a window of ``maximum opportunity'' for its experimental observation. In fact, an information theoretic interpretation can be offered: the log-likelihood is the sum of the information content of the Higgs decays, defined by $\sum_{k=1}^m-\ln(p_k)$, which is maximized precisely for the observed Higgs mass. An estimate of the Higgs mass based on the maximum of the photons branching ratio has also been presented in~\cite{hmaxphotons}.
%
\section{MEP inferences on a new physics scenario}
\label{sec2}

Several beyond the Standard Model scenarios add particles the Higgs boson might decay into. Any channel representing a new Higgs decay might potentially move the Higgs mass that maximizes its entropy. In this respect, imposing that $\hat{M}_H$ remains unaltered, or within certain confidence region around the measured Higgs mass, might narrow the parameters space of the model. Consider then a new channel whose partial width is $\Gamma_{\chi\chi}$, a function of $M_H$, $m_\chi$ the mass of the new particle, and $\lambda$ its coupling to the Higgs boson. 

The branching ratios after including the contribution of the new channel are given by 
\begin{equation}
\tilde{p}_k =\frac{p_k}{1+p_{\chi\chi}}\;\; ,\;\; \tilde{p}_{\chi\chi}=\frac{p_{\chi\chi}}{1+p_{\chi\chi}}
\label{a6}
\end{equation}
where $p_k = \Gamma_k/\Gamma_{SM}$ and $p_{\chi\chi}=\Gamma_{\chi\chi}/\Gamma_{SM}$.

Now, the mass dependent part of the entropy of the Higgs decays is calculated with these new branching ratios
\begin{eqnarray}
 \tilde{S}_\infty(M_H|\pmb{\theta}_{SM},m_\chi,\lambda) &=& \ln\left(\tilde{p}_{\chi\chi}\prod_{k=1}^m\tilde{p}_k\right)\label{a7} \\ 
= S_\infty(M_H|\pmb{\theta}_{SM})\nonumber &+& \ln\left(\frac{p_{\chi\chi}}{(1+p_{\chi\chi})^{m+1}}\right)
\end{eqnarray}
where $S_\infty(M_H|\pmb{\theta}_{SM})$ is given by Eq.~(\ref{a4b}).

Deriving both sides of this equation and requiring that $\frac{d\tilde{S}}{dM_H}=0$ give us the following differential equation
\begin{equation}
\frac{dS_\infty}{dM_H}+\left(\frac{1}{p_{\chi\chi}}-\frac{m+1}{1+p_{\chi\chi}}\right)\frac{dp_{\chi\chi}}{dM_H}=0
\label{a8}
\end{equation}

Now we can seek for solutions of this equation in a mass region around the experimental Higgs mass for a given point in the plane $m_\chi$ {\it versus} $\lambda$.  

Let us present now an specific example of application of MEP inference in a new physics scenario.  A class of interesting models are those where the Higgs boson gains an extra invisible mode, in special, those ones where dark matter couples to the Higgs field as the Higgs-portal models~\cite{higgsportals}. In Ref.~\cite{scalardm}, several experimental constraints are used to determine a viable portion of the parameters space for a real scalar dark matter $\chi$ interacting with the SM Higgs boson doublet $H$ according to ${\cal L}_{int} = \frac{\lambda}{2}\chi^2H^\dagger H$. The dark matter mass $m_\chi$ is constrained to lie in a small region of $55$--$62$ GeV, and a coupling $\lambda$ in the range $10^{-2}$--$10^{-3}$ in order to evade direct detection and relic abundance bounds plus the LHC bound on an invisible Higgs decay ($\tilde{p}_{\chi\chi}=BR_{inv}<0.19$). In order to place constraints on the model, we mapped the allowed region found in Ref.~\cite{scalardm} and searched for $\hat{M}_H$ including the new dark matter decay channel $p_{14}=\tilde{p}_{\chi\chi}=BR(H\rightarrow \chi\chi)$. 
\begin{center}
\begin{figure}
\includegraphics[scale=0.75]{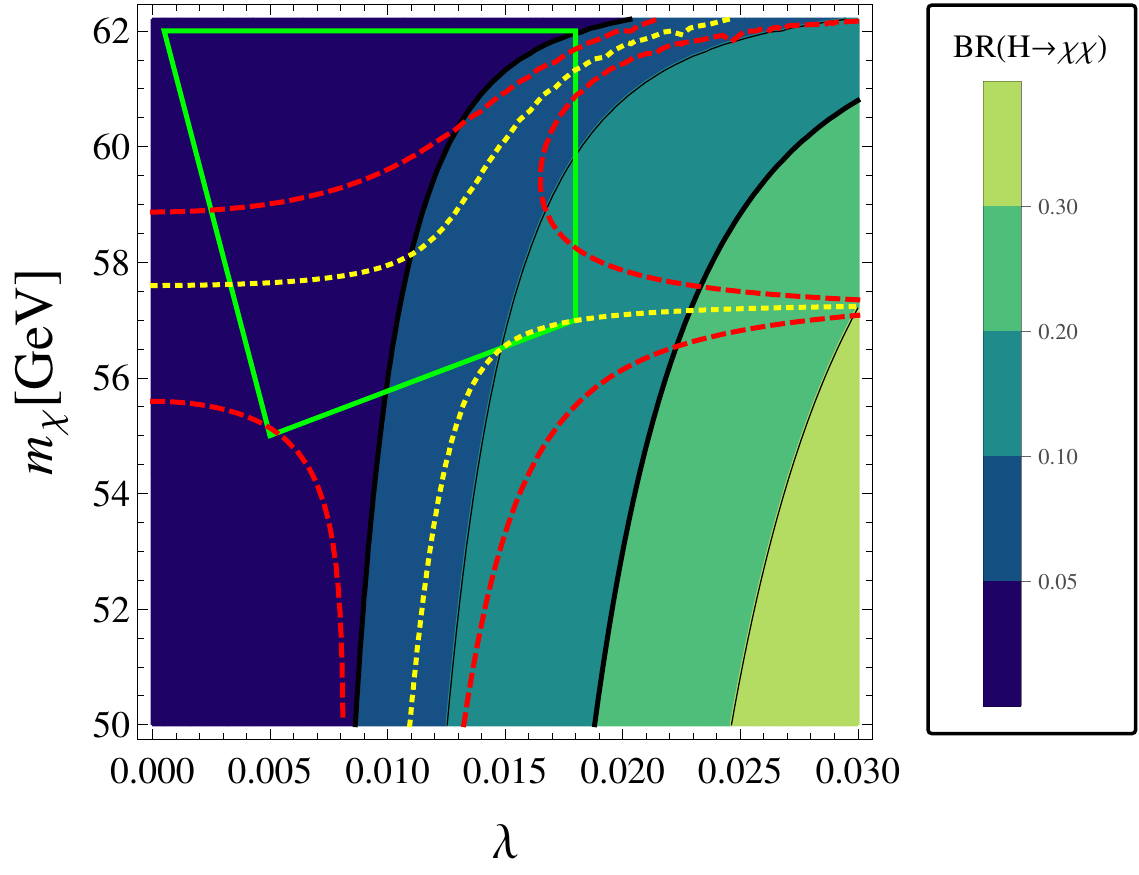}
\caption{The allowed region by direct and indirect (relic density) as found in Ref.~\cite{scalardm} is inside the green quadrilateral. Also shown are the isocontours of the Higgs branching ratio to dark matter. The points inside the region between the dashed red lines are those which maximize the Higgs decays entropy within the 95\% CL experimental region. The yellow dotted lines represent the solution to the Eq.~(\ref{a8}) for the central Higgs mass of $125.14$ GeV.}
\label{fig3}
\end{figure}
\end{center}

In Fig.~(\ref{fig3}) we show the region allowed by direct detection and relic density measurements inside the solid green quadrilateral~\cite{scalardm}, the shaded areas represent the Higgs branching ratio to scalar dark matter in this model, and region between the dashed red curves are those points respecting $\hat{M}_H\in [124.67,125.61]$ GeV, the 95\% CL region of our combined LHC results of the Higgs mass. The yellow dotted lines represent the solution to the Eq.~(\ref{a8}) for the central Higgs mass of $125.14$ GeV.

This result can be seen as prediction of MEP actually. If an invisible decay channel is found in the Run2 of LHC, for example, and an scalar dark matter is confirmed, the experimental result on the coupling to the Higgs and the dark matter mass can be compared to MEP inference. Moreover, as the Higgs mass measurement gets more accurate, the region between lines will shrink to a narrower area leading to a more sharply prediction. On the other, if MEP is correct, no scalar dark matter from a Higgs portal can exist inside the green quadrilateral but outside the dashed lines. We see, by the way, that nearly half the current allowed region would be already ruled out by MEP.

\section{Entropy production in Higgs decays to massless quanta}
\label{sec3}

From the thermodynamic point of view, each decay of the Higgs boson increases the entropy of the bath consisting of SM particles as the energy concentrated in the Higgs boson mass is dissipated into an increased number of quanta. In this respect, photons and gluons can be considered the more efficient channels to increase entropy once they are massless particles~\cite{science, bekenstein}. In Fig.~(\ref{fig1}) we show $P_\gamma(\hat{M}_H)$, obtained exactly as $P_H$, but maximizing the Gibbs-Shannon entropy of the photons channel decay $S_\gamma$. The maximum points of both PDFs are very close.  In fact, the gluonic peak is also near, $\sim 119$ GeV, which lead us to hypothesize that the massless channels play an important role in the explanation of why $S_H$ displays a maximum in first place. As a function of the Higgs mass, $S_H$ is small wherever the Higgs decays to photons and gluons are negligible and strongly increases around the peaks of $p_{\gamma\gamma}=BR(H\rightarrow \gamma\gamma)$ and $p_{gg}=BR(H\rightarrow gg)$. The creation of massive quanta in a co-moving volume of space is less effective in increasing entropy as they organize part of the released energy in their own masses. This is, in part, why entropy is taken as the photon density when we consider the thermodynamic evolution of the Universe: radiation is the best way to spread energy across the space and increase entropy. 

To confirm these intuitions, we calculated $\ln\left(\prod_{k=1}^{11} p_k\right)$ taking into account all channels but the massless ones (photons and gluons) multiplying the remaining branching ratios by $1/(1-p_{\gamma\gamma}-p_{gg})$. The maximum point of the total entropy barely changes to $\hat{M}_H=125.02$ GeV, a negligible shift compared to the value obtained with the full product. 

Withdrawing just the gluonic decay channel, consistently changing the remaining ones, not only marginally changes the point of maximum, now at $\hat{M}_H=125.00$ GeV, but it shortens the distance to the point of maximum of the entropy created in photons decays which now is given by $124.85$ GeV. 

From the observations made above we conclude that: (1) an experimentally compatible inference on the Higgs mass can be made considering only tree-level Higgs couplings; (2) entropy creation from decays to massless  gauge bosons, especially the photon, is maximized close to the same Higgs mass of the full collectively determined entropy; (3) if photons were the only massless fields the Higgs could decay into, the agreement of the maximal points would  improve considerably. 

All these findings suggest that the decay of a system of scalar particles participating an electroweak symmetry breaking mechanism, as in SM, respects MEP concerning the scalars mass in two independent ways: collectively including all particles the Higgs couples to generate masses and individually for the massless gauge bosons related to the unbroken symmetries. Nevertheless, the interplay between massive and massless modes is necessary to accurately determine the Higgs mass.
\section{Discussions and perspectives}
\label{sec4}

The reason why the massless modes peak around the measured Higgs mass might be due the interplay of all the other SM parameters and perhaps by the structure of its symmetry group and the form of the Higgs sector~\cite{hmaxphotons}. Different parameters, for example, lead to another Higgs mass according to the MEP constraint. Removing a channel of the product of branching ratios may perturb $\hat{M}_H$. In particular, the light fermion decays, whose observation is beyond the LHC capabilities, cannot be neglected and the inference $\hat{M}_H=125.04$ GeV would change a lot if they were withdrawn. 

From the point of view of the information theory, MEP inference applies to any new model affecting the Higgs boson decays apart from physical considerations. Its scope of applications is wide, for example, extended Higgs sectors, beyond the Standard Model scenarios predicting new SM Higgs decays and/or deviations of SM Higgs couplings, models for electroweak baryogenesis and Higgsogenesis, Higgs-inflaton proposals and new physics solutions to absolute stabilization of the vacuum. By the way, it is worthy mentioning that MEP should apply to any new Higgs boson from a new physics scenario. Requiring maximization of entropy might bring information about decay rates and its mass as we have delineated in this work.

Testing the validity of MEP in Higgs decays is also straightforward. Any new information that could be related to the Higgs sector can be checked to keep the maximum of $S_H$ compatible with the measured Higgs mass. We expect that as the experimental errors on the SM parameters get smaller, the gaussian PDFs of Fig.~(\ref{fig1}) get narrower with aligned means.

We conclude this Letter stating that the Maximum Entropy Principle seems to be an innate inference tool for a system composed of a large number of Higgs bosons undergoing a collective decay into SM particles. According to this interpretation, such a system can be placed amongst the most fundamental ones where MEP applies.


\vskip0.5cm
\textbf{Acknowledgments} 

This research was  partially supported by the Conselho Nacional de Desenvolvimento Cient\'{\i}fico e Tecnol\'ogico (CNPq), by grants 303094/2013-3 (A.G.D.) and  11862/2012-8 (R.S.). The work of A.A. and A.G.D. was  supported by Funda\c{c}\~{a}o de Amparo \`{a} Pesquisa do Estado de S\~ao Paulo (FAPESP), under the grant  2013/22079-8. We also want to thank Adriano Natale, Yan Levin and David D'Enteria for helpful discussions and suggestions. 






\end{document}